# Torsional and SH surface waves in an isotropic and homogenous elastic half-space characterized by the Toupin-Mindlin gradient theory

**P.A. Gourgiotis** [1]* and **H.G. Georgiadis** [2]

[1] *Department of Mechanical and Structural Engineering, University of Trento, Trento, I-38123, Italy*

[2] *Mechanics Division, National Technical University of Athens, Zographou, GR-15773, Greece*

**Abstract.** The existence of torsional and SH surface waves in a half-space of a homogeneous and isotropic material is shown to be possible in the context of the complete Toupin-Mindlin theory of gradient elasticity. This finding is in marked contrast with the well-known result of the classical theory, where such waves do not exist in a homogeneous (isotropic or anisotropic) half-space. In the context of the classical theory, this weakness is usually circumvented by modeling the half-space as a layered structure or as having non-homogeneous properties. On the other hand, employing a simplified version of gradient elasticity (including only one microstructural parameter and an additional surface-energy term), Vardoulakis and Georgiadis (1997), and Georgiadis et al. (2000), showed that such surface waves may exist in a homogeneous half-space only if a certain type of gradient anisotropy is included in the formulation. On the contrary, in the present work, we prove that the complete Toupin-Mindlin theory of isotropic gradient elasticity (with five microstructural parameters) is capable of predicting torsional and SH surface waves in a purely isotropic and homogeneous material. In fact, it is shown that torsional and SH surface waves are dispersive and can propagate at any frequency (i.e. no cut-off frequencies appear). The character of the dispersion (either normal or anomalous) depends strongly upon the microstructural characteristics.

* Corresponding author: Tel.: +39 0461 282594; Fax: +39 0461 282599.
 E-mail address: p.gourgiotis@unitn.it (Panos A. Gourgiotis)

## 1. Introduction

It is well known that the classical theory of elasticity fails to predict the existence of torsional and antiplane SH (horizontally polarized) surface waves in a *homogeneous* (isotropic or anisotropic) half-space with a free surface (Rayleigh, 1885). On the contrary, both plane stress/strain and axisymmetric surface waves of the Rayleigh type are predicted by the classical theory. Moreover, surface waves of the shear type (i.e. torsional and SH) are known to exist in nature. Indeed, this type of waves has been observed in the context of both non-destructive testing (Kraut, 1971) and seismology (Bullen and Bolt, 1985). In fact, torsional and SH surface waves are the most destructive waves in an earthquake and they can propagate for very long distances without much loss of energy.

As was pointed out by Vardoulakis and Georgiadis (1997), the situation concerning the non-existence of SH and torsional surface waves in a homogeneous half-space, within the context of classical linear elasticity, is translated mathematically to the violation of the pertinent *complementing* (or *consistency*) condition in a semi-infinite domain for the system consisting of the scalar Helmholtz partial differential equation (governing antiplane and torsional motions), a zero Neumann boundary condition for the traction-free surface, and a finiteness condition at infinity. Thompson (1969), showed that the complementing condition implies that all surface waves propagate with non-zero velocity. This condition is obviously satisfied when both dilatational and shear deformations are allowed to take place in the half-space, and thus, Rayleigh surface waves are predicted by the classical theory in the cases of plane stress/strain and general axisymmetric motions. However, the complementing condition is not satisfied in the cases of torsional and antiplane shear (SH) motions, and thus, the corresponding types of waves cannot be predicted by the conventional theory.

In the context of the classical theory of elasticity this drawback is circumvented usually by modeling the half-space as a *layered* structure (Love waves) or as having *non-homogeneous* properties. Regarding the existence and propagation of SH surface waves, a thorough review up to the late 1980s was given by Maugin (1988), whereas an interesting more recent study on the propagation of antiplane SH surface waves (Bleustein – Gulyaev) in a functionally graded material is due to Collet et al. (2006). In addition, Shuvalov et al. (2009) investigated the propagation of SH surface waves in an anisotropic periodic half-spaces in which the material properties within each period are arbitrary. Achenbach and Balogun (2010) examined a purely elastic half-space whose shear modulus and mass density depend arbitrarily on the depth and gave a general solution at high frequencies. In the same context, Ting (2010) investigated the propagation of SH surface waves in a monoclinic half-space with variable density and elastic moduli, and obtained an asymptotic solution for large wavenumbers. Finally, Du and Su (2013) investigated the propagation of SH surface waves in a stochastically homogeneous half-space (with random density in the depth direction) and found interesting dispersion and attenuation properties. On the other hand, regarding the propagation of torsional surface waves the literature is rather limited. Mention should be made of the early work by Meissner (1921), who showed that in an inhomogeneous elastic half-space with quadratic variation of shear modulus and density varying linearly with depth, torsional surface waves do exist. Later, Vardoulakis (1984) showed that the same is true for a Gibson half-space, i.e. for a half-space with shear modulus varying linearly with depth and with constant density. The possibility of surface torsional waves in an elastic half-space with void pores has been examined by Dey et al. (1993), where it was shown that such a half-space can allow two types of torsional surface waves, both being dispersive. More recently, Chattaraj et al. (2011) studied the propagation of torsional surface waves in a poroelastic layer lying over an inhomogeneous elastic half-space under initial stress.

In the context of gradient theories, SH and torsional surface waves have been examined by Vardoulakis and Georgiadis (1997), and Georgiadis et al. (2000), respectively. In these works, a *simplified* version of gradient elasticity with *surface energy* was employed involving two additional material constants (besides the standard two Lamé moduli): the so-called gradient coefficient $c$ and a material parameter $b$ accounting for gradient anisotropy. It was shown that including surface-energy terms (i.e. gradient anisotropy) is necessary for predicting SH and torsional surface waves. Indeed, the standard simplified version of gradient isotropic elasticity (i.e. without surface energy), utilizing a *single* gradient material constant, although greatly facilitates the analysis of boundary value problems (see e.g. Georgiadis et al., 2004; Gourgiotis and Georgiadis, 2009; Gao and Ma, 2009; Giannakopoulos et al. (2012) and references therein), is not capable of predicting these types of surface waves in a homogeneous and isotropic half-space. An analogous situation is encountered in the nonlocal integral-type elasticity theory, which is also incapable of explaining the occurrence of such waves (Eringen, 1972).

In the present work, we employ the *complete* Toupin-Mindlin theory of gradient elasticity with micro-inertia (Toupin, 1962; Mindlin, 1964) and show that this theory is capable of predicting SH and torsional surface waves in a purely *isotropic* and *homogeneous* half-space. Indeed, contrary to the works by Vardoulakis and Georgiadis (1997), and Georgiadis et al. (2000), where a simplified version of the theory was used, neither anisotropy in the material response nor any surface-energy term is needed in the formulation, for the prediction of such waves. In the complete Toupin-Mindlin theory, the full constitutive relations in the isotropic case involve *five* microstructural parameters (these constants are additional to the standard Lamé constants to characterize the material response), providing thus a more detailed modeling of microstructured materials as compared to the simplified version (including only one additional material parameter) or other generalized continuum theories - like the standard couple-stress theory - employed in the past for examining wave propagation problems (e.g. Georgiadis and Velgaki, 2003; Georgiadis et al., 2004; Vavva et al., 2009; Gourgiotis et al., 2013; Rosi et al., 2014; Piccolroaz and Movchan, 2014; Morini et al., 2014). In the present formulation, a *micro-inertia* term is also included, since previous experience with gradient analyses of surface waves showed that this term is indeed important at high frequencies (Georgiadis et al., 2004; Filopoulos et al. 2010; Gourgiotis et al., 2013). The inclusion of the micro-inertia term leads to an explicit appearance of the intrinsic material length $2h$, which, in turn, can be associated with the material microstructure. Recently, Polyzos and Fotiadis (2012), using a simple one-dimensional lattice model of one-neighbor interaction reproduced the field equations of Toupin-Mindlin theory and correlated the internal lengths parameters with the actual microstructure of the material. Moreover, Shodja et al. (2013) utilizing ab initio DFT calculations evaluated the characteristic material lengths of the Toupin-Mindlin theory for several fcc and bcc metal crystals.

The contents of our paper are as follows: In Section 2, we summarize the basic dynamical equations of the Form II of Toupin-Mindlin gradient theory and examine the effects of strain-gradients in the propagation of plane waves in an infinite medium. It is worth noting that, unlike the case of classical theory, in gradient elasticity both dilatational and distortional waves become dispersive. In addition, the conditions for positive definiteness of the strain-energy density are provided in the context of the complete isotropic Toupin-Mindlin theory. Next, in Section 3, we investigate the propagation of torsional surface waves in a homogeneous and isotropic gradient half-space. The solution is derived with the use of Hankel transforms. A parametric analysis of the pertinent dispersion equation reveals the conditions for the existence of such waves. In Section 4, free time-harmonic SH motions are considered for a homogeneous and isotropic gradient half-space. The analysis is based on the Fourier transform and on a parametric study of the resulting dispersion equation. Numerical results and asymptotic estimates regarding the dispersion characteristics of torsional and SH waves are presented in Section 5. The dependence of the phase and group velocities upon the wavenumber and the microstructural characteristics of the material is studied in detail. It is shown that, for a material with a positive definite strain-energy density, torsional and SH surface waves can propagate at *any* frequency (i.e. no cut-off frequencies appear). Moreover, the character of the dispersion (either normal or anomalous) depends upon the size of the material microstructure.

Our results can be useful in wave-propagation studies (e.g. in relation with non-destructive testing and evaluation) for granular materials such as ceramics, composites, foams, masonry structures, bone tissues, glassy and semi-crystalline polymers, where their macroscopic behavior is strongly influenced by the microstructural characteristic lengths especially at high frequencies or in the presence of large stress (or strain) gradients.

## 2. Fundamentals of strain gradient elastodynamics

In this Section, we briefly present the basic elastodynamic equations of the isotropic Toupin-Mindlin theory of strain-gradient elasticity. A detailed presentation of the theory including inertial and micro-inertial effects can be found in the fundamental paper of Mindlin (1964) and the recent papers by Georgiadis et al. (2004), and Gourgiotis et al. (2013). According to this theory, each material particle has three degrees of freedom (the displacement components – just as in the classical theory) and the micro-density does not differ from the macro-density. Also, first-order gradient terms of strain and velocity, in addition to the classical (i.e. zero-order gradient) terms, are included in the strain and the kinetic energy densities, respectively.

For a continuum with microstructure fully composed of sub-particles (micro-media) having the form of unit cells (cubes), the following expression of the kinetic-energy density $T$ is obtained with respect to a Cartesian coordinate system $Ox_1x_2x_3$ (Mindlin, 1964)

$$T = \frac{1}{2}\rho \dot{u}_p \dot{u}_p + \frac{1}{6}\rho h^2 \left(\partial_p \dot{u}_q\right)\left(\partial_p \dot{u}_q\right), \qquad (1)$$

where $\rho$ is the mass density, $2h$ is the size of the cube edges of the unit cell, $u_p$ is the displacement vector, $\partial_p(\ ) \equiv \partial(\ )/\partial x_p$, the superposed dot denotes time derivative, and the Latin indices span the range (1,2,3) (indicial notation and summation convention is used throughout).

The second term in the RHS of Eq. (1), involving the velocity gradients, represents the *micro-inertia* of the continuum. This term, which is not encountered within classical continuum mechanics, reflects the more detailed description of motion in the present theory.

Also, the strain-energy density function for a linear and isotropic continuum assumes the following form (Mindlin, 1964)

$$W = \frac{1}{2}\lambda \varepsilon_{pp}\varepsilon_{qq} + \mu \varepsilon_{pq}\varepsilon_{pq} + a_1 \kappa_{ppj}\kappa_{jqq} + a_2 \kappa_{jpp}\kappa_{jqq}$$
$$+ a_3 \kappa_{ppj}\kappa_{qqj} + a_4 \kappa_{jpq}\kappa_{jpq} + a_5 \kappa_{jpq}\kappa_{qpj} \;, \tag{2}$$

where $\varepsilon_{pq} = (1/2)\left(\partial_p u_q + \partial_q u_p\right) = \varepsilon_{qp}$ is the linear strain tensor, and $\kappa_{rpq} = \kappa_{rqp} = \partial_r \varepsilon_{pq}$ is the strain gradient (third order) tensor. This is Form II in Mindlin's (1964) paper. In addition, $(\lambda, \mu)$ are the standard Lamé constants and $a_q$ ($q = 1,...,5$) are the five additional material constants having dimensions of [force]. It is worth noting that the frequently used simplified version of gradient elasticity is obtained from Eq. (2) by setting: $a_2 = \lambda c/2$, $a_4 = \mu c$, and $a_1 = a_3 = a_5 = 0$ (see e.g. Georgiadis et al., 2004; Gao and Ma, 2009; Vavva et al., 2009).

In view of Eq. (2), the constitutive equations become

$$\tau_{pq} = \frac{\partial W}{\partial \varepsilon_{pq}} = \lambda \delta_{pq}\varepsilon_{jj} + 2\mu \varepsilon_{pq} \;, \tag{3}$$

$$m_{rpq} = \frac{\partial W}{\partial(\partial_r \varepsilon_{pq})} = \frac{1}{2}a_1\left(\delta_{rp}\kappa_{qjj} + 2\delta_{pq}\kappa_{jjr} + \delta_{qr}\kappa_{pjj}\right) + 2a_2 \delta_{pq}\kappa_{rjj}$$
$$+ a_3\left(\delta_{rp}\kappa_{jjq} + \delta_{rq}\kappa_{jjp}\right) + 2a_4 \kappa_{rpq} + a_5\left(\kappa_{qrp} + \kappa_{pqr}\right) \;, \tag{4}$$

where $\delta_{pq}$ is the Kronecker delta, $\tau_{pq}$ is the monopolar stress tensor, and $m_{rpq}$ is the dipolar (or double) stress tensor (a third-rank tensor) expressed in dimensions of [force][length]$^{-1}$. The dipolar stress tensor follows from the notion of dipolar forces, which are anti-parallel forces acting between the micro-media contained in the continuum with microstructure. According to Eqs. (3) and (4), the following symmetries for the monopolar and dipolar stress tensors are noticed: $\tau_{pq} = \tau_{qp}$ and $m_{rpq} = m_{rqp}$.

The equations of motion and the pertinent boundary conditions can be obtained from Hamilton's principle and variational considerations using Eqs. (1) and (2). Indeed, assuming the absence of body forces, the variational form of Hamilton's principle becomes

$$\int_{t_1}^{t_2}\int_V \delta W \, dV \, dt - \int_{t_1}^{t_2}\int_V \delta T \, dV \, dt = \int_{t_1}^{t_2}\left\{\int_S t_q^{(n)} \delta u_q \, dS + \int_S T_{qr}^{(n)} \partial_q (\delta u_r) \, dS\right\} dt, \tag{5}$$

where $V$ is the region occupied by the body, and $S$ is the surface of the body. The symbol $\delta$ denotes weak variations and it acts on the quantity existing on its right. Also, $t_1$ and $t_2$ are two arbitrary instants of time for which the variations $\delta u_q$ are zero at all points of the body. In addition, $t_q^{(n)}$ is the true monopolar traction, $T_{pq}^{(n)}$ is the true dipolar traction, and $n_p$ is the outward unit normal to the boundary along a Section inside the body or along the surface of it (Bleustein, 1967). Examples of the dipolar tractions $T_{pq}^{(n)}$ can be found in the work by Georgiadis and Anagnostou (2008).

The local form of the equations of motion and the traction boundary conditions along a *smooth* boundary assume then the following form (Mindlin, 1964)

$$\partial_p\left(\tau_{pq} - \partial_r m_{rpq}\right) = \rho \ddot{u}_q - \frac{\rho h^2}{3}\left(\partial_{pp}\ddot{u}_q\right) \quad \text{in} \quad V, \tag{6}$$

$$P_q^{(n)} = n_p\left(\tau_{pq} - \partial_r m_{rpq}\right) - D_p\left(n_r m_{rpq}\right) + \left(D_j n_j\right) n_r n_p m_{rpq} + \frac{\rho h^2}{3} n_p\left(\partial_p \ddot{u}_q\right) \quad \text{on} \quad S, \tag{7}$$

$$R_q^{(n)} = n_r n_p m_{rpq} \quad \text{on} \quad S, \tag{8}$$

where $D_p(\ ) \equiv \partial_p(\ ) - n_p D(\ )$ is the surface gradient operator and $D(\ ) \equiv n_p \partial_p(\ )$ is the normal gradient operator. The auxiliary force traction $P_q^{(n)}$ and the auxiliary double force traction $R_q^{(n)}$ are related with the true force traction $t_q^{(n)}$ and the true double force traction $T_{pq}^{(n)}$ through $P_q^{(n)} \equiv t_q^{(n)} + (D_r n_r) n_p T_{pq}^{(n)} - D_p T_{pq}^{(n)}$ and $R_q^{(n)} \equiv n_p T_{pq}^{(n)}$ (Bleustein, 1967). It should be noted that in the case in which edges appear along the boundary, an additional boundary condition should also be imposed (Mindlin, 1964). Moreover, the pertinent kinematical boundary conditions of the theory were derived by Mindlin (1964) (see also, Grentzelou and Georgiadis, 2008), but are omitted here since these are not relevant to our specific problem.

The following point now deserves attention: In the general inertial case, the existence of last term in the LHS of (7) violates the assumption of objective tractions. However, in the quasi-static case and also in the time-harmonic inertial case considered here this difficulty is eliminated. Moreover, as Jaunzemis (1967) pointed out, the difficulty with satisfying objectivity in multipolar theories can circumvented by introducing an effective body force as the difference between the standard body force and the micro-inertia term, and by further assuming that this effective body force is objective, although its constituents are not (see also Georgiadis et al., 2004).

In summary, Equations (3), (4) and (6)-(8) are the governing equations for the isotropic linear gradient elastodynamic theory. Combining Eqs. (3) and (4) with (6), one obtains the equations of motion in terms of displacement components (Mindlin, 1964)

$$(\lambda + 2\mu)(1 - \ell_1^2 \nabla^2) \nabla (\nabla \cdot \mathbf{u}) - \mu (1 - \ell_2^2 \nabla^2) \nabla \times \nabla \times \mathbf{u} = \rho \ddot{\mathbf{u}} - I \nabla^2 \ddot{\mathbf{u}} ,  \qquad (9)$$

where $\nabla^2 ( \ )$ is the Laplace operator, $I = \rho h^2 / 3$ is the micro-inertia coefficient, and $(\ell_1^2, \ell_2^2) > 0$ are the characteristic lengths, defined as

$$\ell_1^2 = \frac{2(a_1 + a_2 + a_3 + a_4 + a_5)}{(\lambda + 2\mu)} > 0 , \quad \ell_2^2 = \frac{(a_3 + 2a_4 + a_5)}{2\mu} > 0 . \qquad (10)$$

In the limit $(\ell_1, \ell_2) \to 0$ and $h \to 0$, the Navier-Cauchy equations of classical dynamic linear isotropic elasticity are recovered from Eqs. (9). The fact that the gradient coefficients $(\ell_1^2, \ell_2^2)$ multiply the higher-order term reveals the *singular-perturbation* character of the gradient theory and the emergence of associated *boundary-layer* effects.

Next, by taking the divergence and curl of Eqs. (9), we obtain the relations governing the propagation of dilatation and rotation, respectively

$$c_p^2 (1 - \ell_1^2 \nabla^2) \nabla^2 (\nabla \cdot \mathbf{u}) = \left(1 - \frac{h^2}{3} \nabla^2\right) \nabla \cdot \ddot{\mathbf{u}} , \qquad (11)$$

$$c_s^2\left(1-\ell_2^2\nabla^2\right)\nabla^2\left(\nabla\times\mathbf{u}\right)=\left(1-\frac{h^2}{3}\nabla^2\right)\nabla\times\ddot{\mathbf{u}}\ ,\tag{12}$$

where $c_p=\left[(\lambda+2\mu)/\rho\right]^{1/2}$ and $c_s=(\mu/\rho)^{1/2}$ are the velocities of the pressure (P) and shear (S) waves, respectively, in the classical (i.e. non-gradient) elasticity theory. Moreover, we note that unlike the corresponding case of classical elastodynamics, the PDEs (11) and (12) are of the fourth order. This implies that wave signals emitted from a disturbance point propagate at different velocities. The last statement can easily be supported by considering time-harmonic plane wave solutions and determining the pertinent dispersion relations. To this end, we consider a plane wave solution in the following form:

$$\mathbf{u}=A\mathbf{d}\exp\left[i\left(\xi(\mathbf{n}\cdot\mathbf{x})-\omega t\right)\right]\ ,\tag{13}$$

where $A$ denotes the amplitude, $(\mathbf{d},\mathbf{n})$ are unit vectors defining the directions of motion and propagation, respectively, $\mathbf{x}$ is the position vector, $\xi$ is the wavenumber, $\omega$ is the circular frequency of the plane wave (taken to be a real quantity), $V=\omega/\xi$ is the phase velocity, and $i^2=-1$. Then, on substituting the solution (13) into Eqs. (11) and (12), we obtain the following relations for the phase velocities of the pressure and shear waves in gradient elasticity

$$V_p=c_p\left(1+\ell_1^2\xi^2\right)^{1/2}\left(1+\frac{h^2}{3}\xi^2\right)^{-1/2},\quad V_s=c_s\left(1+\ell_2^2\xi^2\right)^{1/2}\left(1+\frac{h^2}{3}\xi^2\right)^{-1/2}.\tag{14}$$

Equations (14) show that the propagation velocities of these waves depend on the respective wavenumber. Hence, both waves are dispersive in dipolar gradient elasticity. This finding is in contrast with the result of the standard couple-stress elasticity, where only the shear waves become dispersive (Toupin, 1962). A recent interesting investigation regarding the capability of various gradient and nonlocal type theories to predict the dispersive behavior of traveling waves in comparison with the Born–Karman model of lattice dynamics was given by Fafalis et al. (2012).

To investigate further upon the nature of the dispersion relations in gradient elasticity, we consider the group velocity $V^g = d\omega/d\xi$ at which the energy propagates in a dispersive medium (Achenbach, 1984). In particular, according to Eqs. (14), we obtain

$$V_p^g = V_p + c_p \left( \ell_1^2 - \frac{h^2}{3} \right) \xi^2 \left( 1 + \ell_1^2 \xi^2 \right)^{-1/2} \left( 1 + \frac{h^2}{3} \xi^2 \right)^{-3/2}, \tag{15}$$

$$V_s^g = V_s + c_s \left( \ell_2^2 - \frac{h^2}{3} \right) \xi^2 \left( 1 + \ell_2^2 \xi^2 \right)^{-1/2} \left( 1 + \frac{h^2}{3} \xi^2 \right)^{-3/2}. \tag{16}$$

The following three cases are then distinguished: (i) For $\ell_{1,2}^2 < h^2/3$, Eqs. (15) and (16) imply that $V_{p,s}^g < V_{p,s}$ and thus the dispersion is normal. (ii) For $\ell_{1,2}^2 > h^2/3$, we have $V_{p,s}^g > V_{p,s}$ indicating that the dispersion is anomalous. (iii) For $\ell_{1,2}^2 = h^2/3$ or $(\ell_1, \ell_2, h) \to 0$ (i.e. no material microstructure), the wave velocities degenerate into the non-dispersive velocities of classical elastodynamics.

Finally, the restriction of positive definiteness of the strain energy density $W$ requires the following inequalities for the material constants (a detailed derivation is provided in Appendix A)

$$(3\lambda + 2\mu) > 0, \quad \mu > 0, \tag{17}$$

$$a_4 > 0, \quad a_4 + a_5 > 0, \quad 2a_4 - a_5 > 0,$$

$$b_1 > 0, \quad b_2 > 0, \quad 2b_1 b_2 - 5(a_1 + 4a_2 - 2a_3)^2 > 0, \quad 10a_3 + 6a_4 + a_5 > 0, \tag{18}$$

where

$$b_1 = -4a_1 + 8a_2 + 2a_3 + 6a_4 - 3a_5 \quad \text{and} \quad b_2 = 5(a_1 + a_2 + a_3) + 3(a_4 + a_5). \tag{19}$$

## 3. Torsional surface waves in a gradient elastic half-space

*3.1. Governing equations for time-harmonic torsional motions*

Attention now is directed to the torsional dynamic motions in a gradient-elastic half-space. A torsional motion is one that involves only the circumferential displacement, which is independent of the azimuthal angle. With respect to a system of cylindrical coordinates $(r,\theta,z)$ having unit base vectors $(\mathbf{e}_r,\mathbf{e}_\theta,\mathbf{e}_z)$, the half-space occupies the region $(0 \leq r < \infty, z \geq 0)$ (see Fig. 1).

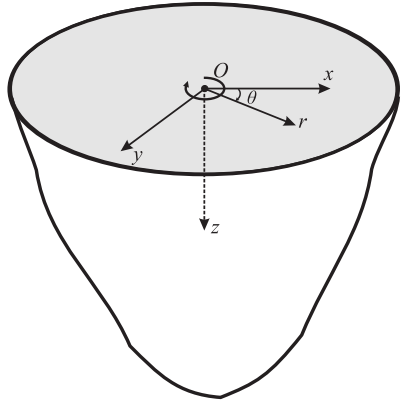

**Fig.1**: An elastic half-space in a state of torsional motions.

In this case, the displacement field assumes the following general form

$$u_r = u_z = 0 , \quad u_\theta = u(r,z,t) \neq 0 , \tag{20}$$

whereas, the non-vanishing components of the strain and strain-gradient tensors are

$$\varepsilon_{r\theta} = \frac{1}{2}\left(\frac{\partial u}{\partial r} - \frac{u}{r}\right) , \quad \varepsilon_{\theta z} = \frac{1}{2}\frac{\partial u}{\partial z} , \tag{21}$$

$$\kappa_{rr\theta} = \frac{\partial \varepsilon_{r\theta}}{\partial r}, \quad \kappa_{\theta\theta\theta} = -\kappa_{\theta rr} = \frac{2\varepsilon_{r\theta}}{r}, \quad \kappa_{zz\theta} = \frac{\partial \varepsilon_{\theta z}}{\partial z},$$

$$\kappa_{r\theta z} = \frac{\partial \varepsilon_{\theta z}}{\partial r}, \quad \kappa_{zr\theta} = \frac{\partial \varepsilon_{r\theta}}{\partial z}, \quad \kappa_{\theta rz} = -\frac{\varepsilon_{\theta z}}{r}. \tag{22}$$

Further, in view of Eqs. (3), (4) and (22) the monopolar and dipolar stresses become

$$\tau_{r\theta} = 2\mu \varepsilon_{r\theta}, \quad \tau_{\theta z} = 2\mu \varepsilon_{\theta z}, \tag{23}$$

$$m_{rr\theta} = (a_3 + 2a_4 + a_5)\kappa_{rr\theta} + a_5\kappa_{\theta rr} + a_3(\kappa_{zz\theta} + \kappa_{\theta\theta\theta}),$$

$$m_{\theta\theta\theta} = (a_1 + 2a_3)(\kappa_{rr\theta} + \kappa_{\theta\theta\theta} + \kappa_{zz\theta}) + 2(a_4 + a_5)\kappa_{\theta\theta\theta},$$

$$m_{\theta rr} = a_1(\kappa_{rr\theta} + \kappa_{\theta\theta\theta} + \kappa_{zz\theta}) + 2a_4\kappa_{\theta rr} + 2a_5\kappa_{rr\theta},$$

$$m_{zz\theta} = (a_3 + 2a_4 + a_5)\kappa_{zz\theta} + a_3(\kappa_{rr\theta} + \kappa_{\theta\theta\theta}),$$

$$m_{\theta zz} = a_1(\kappa_{rr\theta} + \kappa_{\theta\theta\theta} + \kappa_{zz\theta}) + 2a_5\kappa_{zz\theta},$$

$$m_{r\theta z} = 2a_4\kappa_{r\theta z} + a_5(\kappa_{zr\theta} + \kappa_{\theta rz}),$$

$$m_{\theta rz} = 2a_4\kappa_{\theta rz} + a_5(\kappa_{zr\theta} + \kappa_{r\theta z}),$$

$$m_{zr\theta} = 2a_4\kappa_{zr\theta} + a_5(\kappa_{r\theta z} + \kappa_{\theta rz}). \tag{24}$$

Since in the torsional case only shear motions exist, omitting the terms accounted for dilatational deformation in the equations of motion (Eq. (9)), we obtain

$$\left[\ell^2\left(\nabla^2 - \frac{1}{r^2}\right)^2 - \left(\nabla^2 - \frac{1}{r^2}\right)\right]u = \frac{I}{\mu}\left(\nabla^2 - \frac{1}{r^2}\right)\ddot{u} - \frac{\ddot{u}}{c_s^2}, \tag{25}$$

where $\ell \equiv \ell_2$, and $\nabla^2(\ ) \equiv \partial_r^2(\ ) + r^{-1}\partial_r(\ ) + \partial_z^2(\ )$ is the Laplace operator in cylindrical polar coordinates depending now only upon the variables $(r, z)$. In the absence of gradient effects (i.e. when $\ell = 0$ and $h = 0$), Eq. (25) degenerates into the standard wave equation of the second order governing torsional motions. It is worth noting that the *dispersive* character of torsional waves in gradient elasticity can be immediately inferred by the structure of the differential operator in (25).

In the sequel, a *steady state* is considered where, as is well-known (see e.g. Achenbach, 1984), the displacement varies in the following time-harmonic manner

$$u(r,z,t) = u(r,z) \cdot e^{-i\omega t} \ . \tag{26}$$

The above 'decomposition' reduces Eq. (25) to the form

$$\ell^2 \left( \nabla^2 - \frac{1}{r^2} \right)^2 u - g \left( \nabla^2 - \frac{1}{r^2} \right) u - k^2 u = 0 \ , \tag{27}$$

where $k \equiv k(\omega) = \omega/c_s$ and $g \equiv g(\omega) = 1 - (I\omega^2/\mu)$. In what follows, as is standard in this type of problems, it is implied that all field quantities are to be multiplied by the time-harmonic factor $\exp(-i\omega t)$ and that the real part of the resulting expression is to be taken.

The pertinent boundary conditions for a *traction-free* half-space follow from Eqs. (6). For a boundary defined by the plane $z = 0$ with $\mathbf{n} = (0, 0, -1)$, they assume the following form

$$P_\theta^{(n)}(r, z = 0) = \tau_{\theta z} - \frac{\partial m_{rz\theta}}{\partial r} - \frac{\partial m_{zz\theta}}{\partial z} - \frac{\partial m_{zr\theta}}{\partial r} - \frac{m_{r\theta z}}{r} - \frac{m_{\theta rz}}{r} - \frac{2m_{zr\theta}}{r} - I\omega^2 \frac{\partial u}{\partial z} = 0 \ , \tag{28}$$

$$R_\theta^{(n)}(r, z = 0) = m_{zz\theta} = 0 \ . \tag{29}$$

It should be remarked that in the work of Georgiadis et al. (2000), a simplified boundary condition for the monopolar traction $P_\theta^{(n)}$ was used instead of the exact relation in (28). Next, employing the constitutive equations (23) and (24) in conjunction with the geometric relations (21) and (22), the boundary conditions are written in terms of the displacement component $u(r,z)$ as

$$\left( \ell^2 - \bar{a}_3 \right) \frac{\partial^3 u}{\partial z^3} - \left( 2\ell^2 - \bar{a}_3 \right) \left( \nabla^2 - \frac{1}{r^2} \right) \frac{\partial u}{\partial z} + g \frac{\partial u}{\partial z} = 0 \qquad \text{for } z = 0 \text{ and } 0 \leq r < \infty \ , \tag{30}$$

$$\left( \ell^2 - \bar{a}_3 \right) \frac{\partial^2 u}{\partial z^2} + \bar{a}_3 \left( \nabla^2 - \frac{1}{r^2} \right) u = 0 \qquad \text{for } z = 0 \text{ and } 0 \leq r < \infty \ , \tag{31}$$

with $\bar{a}_3 \equiv a_3/2\mu$. The material constant $\bar{a}_3$ has dimensions of $[\text{length}]^2$ and can take positive or negative values. In particular, from the requirement of positive definiteness of the strain-energy density, the inequality $|\bar{a}_3| < \ell^2$ should always be satisfied (see Appendix A).

*3.2. Integral-transform analysis and dispersion equation*

In view of the axisymmetry of the problem and in order to suppress the $r$-dependence in the governing equations, the *Hankel transform* of order one is employed (see e.g. Davies, 2002)

$$f^*(\xi,z) = \int_0^\infty f(r,z) J_1(\xi r) \cdot r\, dr, \quad f(r,z) = \int_0^\infty f^*(\xi,z) J_1(\xi r) \cdot \xi\, d\xi, \tag{32}$$

where $J_1(\ )$ is the Bessel function of the first kind and order one. Under the operation of the direct Hankel transform and assuming the required regularity conditions for $u(r,z)$, Eq. (27) is transformed into the following ordinary differential equation

$$\ell^2 \frac{d^4 u^*}{dz^4} - \left(2\ell^2 \xi^2 + g\right) \frac{d^2 u^*}{dz^2} + \left(\ell^2 \xi^4 + g\xi^2 - k^2\right) u^* = 0, \tag{33}$$

where the range of the transform variable $\xi$ can be extended, by analytic continuation, into the whole complex plane. Now, Eq. (33) has the following *bounded* solution as $z \to \infty$

$$u^*(\xi,z) = B(\xi) e^{-\beta z} + C(\xi) e^{-\gamma z} \quad \text{for} \quad z \geq 0, \tag{34}$$

provided that the $\xi$-plane has been cut appropriately, taking the branch cuts for $(\beta,\gamma)$ as shown in Fig. 2. In this case, $B(\xi)$ and $C(\xi)$ are unknown functions of the wavenumber $\xi$ and $(\beta,\gamma)$ are the relevant roots given by

$$\beta \equiv \beta(\xi) = \left(\xi^2 - \sigma^2\right)^{1/2}, \quad \text{with} \quad \sigma = \frac{\left[\left(g^2 + 4\ell^2 k^2\right)^{1/2} - g\right]^{1/2}}{\left(2\ell^2\right)^{1/2}} > 0, \tag{35}$$

$$\gamma \equiv \gamma(\xi) = (\xi^2 + \tau^2)^{1/2}, \quad \text{with} \quad \tau = \frac{\left[(g^2 + 4\ell^2 k^2)^{1/2} + g\right]^{1/2}}{(2\ell^2)^{1/2}} > 0 \ . \tag{36}$$

The criterion for surface waves in this case is that the displacement $u$ decays exponentially with the distance $z$ from the free surface. Such a case for a *homogeneous* half-space is precluded according to the classical elasticity theory but can arise, as it is shown below, within the present isotropic gradient elasticity theory. Indeed, in view of the analysis leading to (34) and taking into account the structure of the inverse Hankel transform (32)$_2$, we now explore the possibility of *progressive-wave* solutions of Eq. (25) having the form of a distinct time harmonic component

$$u(r, z, t; \xi) = u^*(\xi, z) J_1(\xi r) \xi \, e^{-i\omega t} \ , \tag{37}$$

where the propagation wavenumber $\xi$ is taken to be a real quantity, and $(\beta, \gamma)$ defined in (35) and (36) are taken to be *real* and *positive* functions. The latter restriction is satisfied if and only if $\sigma < |\xi|$. Taking a real wavenumber excludes the possibility of localized *standing* waves (i.e. leaky or evanescent motions). Finally, we remark that a general surface-wave motion (synthesis) can be derived by superposition as a Hankel inversion integral (Eringen and Suhubi, 1975)

$$u_s(r, z, t) = \int_0^\infty u^*(\xi, z) J_1(\xi r) \xi \cdot e^{-i\omega(\xi)t} d\xi \ . \tag{38}$$

where integration is over the whole range of wavenumbers. Note that in Eq. (37) the frequency $\omega$ and the wavenumber $\xi$ are related through the pertinent dispersion equation (c.f. (40)) in order for each distinct torsional surface wave component to propagate.

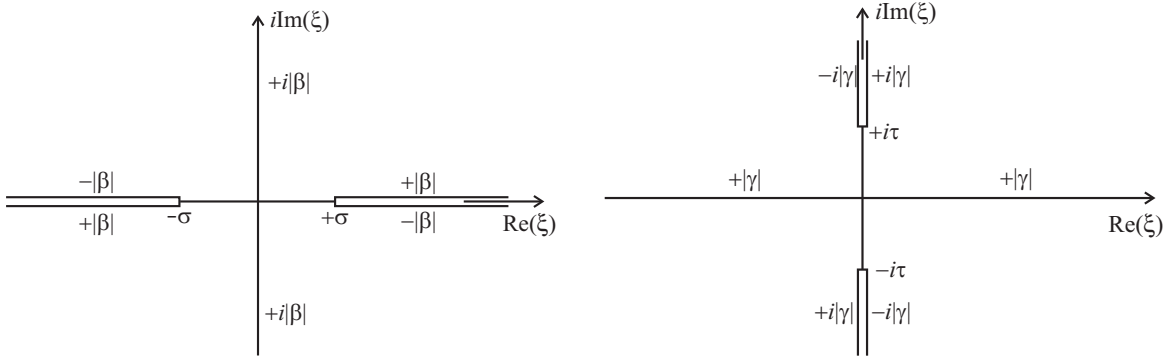

**Fig. 2:** The cut complex $\xi$-plane for the functions $\beta(\xi)$ and $\gamma(\xi)$.

The appropriate *dispersion equation* is now obtained by enforcing the traction-free boundary conditions (30) and (31) along the half-space surface $z=0$. Transforming the boundary conditions, the following linear homogeneous system results for the unknown amplitudes $B$ and $C$

$$\begin{bmatrix} \left(\ell^2\gamma^2 - \bar{a}_3\xi^2\right)\beta & \left(\ell^2\beta^2 - \bar{a}_3\xi^2\right)\gamma \\ \left(\ell^2\beta^2 - \bar{a}_3\xi^2\right) & \left(\ell^2\gamma^2 - \bar{a}_3\xi^2\right) \end{bmatrix} \begin{bmatrix} B \\ C \end{bmatrix} = \begin{bmatrix} 0 \\ 0 \end{bmatrix}, \qquad (39)$$

which has a nontrivial solution if and only if the determinant of the matrix $D(\xi,V)$ is zero

$$D(\xi,V) \equiv \beta\left(\ell^2\gamma^2 - \bar{a}_3\xi^2\right)^2 - \gamma\left(\ell^2\beta^2 - \bar{a}_3\xi^2\right)^2 = 0, \qquad (40)$$

where $V = \omega/\xi$ is the gradient phase velocity of the torsional surface waves (note that, in what follows, $V$ should not be confused with the volume of the body defined in Section 1). Equation (40) is the *dispersion relation* for the motion of progressive torsional surface waves in a gradient-elastic *homogeneous* and *isotropic* half-space. From this equation, dispersion curves are obtained and will be presented in Section 5. Regarding the nature of the dispersion equation (40), the following observations are in order:

(*i*) Torsional surface waves exist if and only if

$$\ell^2 \neq 0 \quad \text{and} \quad \bar{a}_3 \neq 0 ,  \tag{41}$$

the cases, ($\ell^2 = 0$) or ($\ell^2 \neq 0$ and $\bar{a}_3 = 0$) lead to non-existence of such waves. Indeed, the case $\ell^2 = 0$ leads to a degeneracy of the governing PDE (27) and thus the traction-free boundary conditions in (30) and (31) cannot be satisfied. On the other hand, for $\bar{a}_3 = 0$, the dispersion equation degenerates to: $\beta(\xi) = 0$ (since $\gamma(\xi) \neq 0$ and $\beta(\xi) \neq \gamma(\xi)$, for all real wavenumbers $\xi$), which, according to Eq. (39), implies that the amplitude $C$ is zero and therefore, in this case, no torsional surface waves exist. It is worth noting that when $a_3 = 0$, the complete Toupin-Mindlin gradient theory, employed in the present work, degenerates to the simplified version of gradient isotropic elasticity which consequently is unable to predict torsional surface waves (or SH surface waves, as it will be shown in the next section).

(*ii*) The dispersion equation (40) being an irrational algebraic equation is a *monomode* equation and this is in some contrast with the infinity of modes resulting from transcendental equations, which correspond to non-homogeneous models of a half-space supporting torsional surface waves in the classical theory (Meissner, 1921; Vardoulakis, 1984).

**4. SH surface waves in a gradient elastic half-space**

*4.1. Governing equations for time-harmonic antiplane shear motions*

Antiplane shear (i.e. horizontally polarized – SH) motions are now examined in a homogeneous and isotropic gradient-elastic half-space. With respect to an *Oxyz* Cartesian coordinate system, the half-space occupies the region ($-\infty < x < \infty$, $y \geq 0$) and is long enough in the $z$-direction to allow an antiplane shear state when loadings act in the same direction (Fig. 3). In this case any problem is essentially two-dimensional depending on $(x, y)$. Then, the displacement field assumes the following form

$$u_x = u_y = 0 , \quad u_z \equiv w(x, y, t) \neq 0 . \tag{42}$$

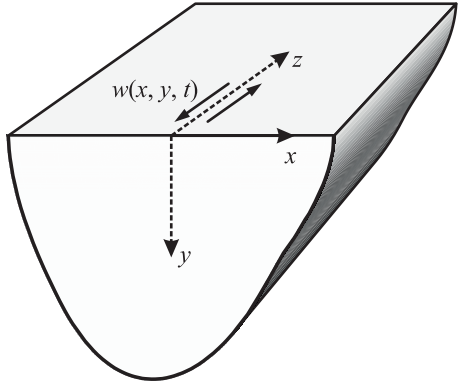

**Fig. 3**: An elastic half-space in an antiplane shear state.

Accordingly, the non-vanishing components of the strain and the strain-gradient tensors are

$$\varepsilon_{xz} = \frac{1}{2}\frac{\partial w}{\partial x}, \quad \varepsilon_{yz} = \frac{1}{2}\frac{\partial w}{\partial y}, \tag{43}$$

$$\kappa_{xxz} = \frac{1}{2}\frac{\partial^2 w}{\partial x^2}, \quad \kappa_{yyz} = \frac{1}{2}\frac{\partial^2 w}{\partial y^2}, \quad \kappa_{xyz} = \kappa_{yxz} = \frac{1}{2}\frac{\partial^2 w}{\partial x \partial y}. \tag{44}$$

In view of the above, and taking into account the constitutive equations in (3) and (4), the monopolar and dipolar stresses become

$$\tau_{xz} = \mu \frac{\partial w}{\partial x}, \quad \tau_{yz} = \mu \frac{\partial w}{\partial y}, \tag{45}$$

$$m_{xxz} = \mu \ell^2 \frac{\partial^2 w}{\partial x^2} + \mu \bar{a}_3 \frac{\partial^2 w}{\partial y^2}, \quad m_{yyz} = \mu \ell^2 \frac{\partial^2 w}{\partial y^2} + \mu \bar{a}_3 \frac{\partial^2 w}{\partial x^2},$$

$$m_{xyz} = m_{yxz} = \mu \left( \ell^2 - \bar{a}_3 \right) \frac{\partial^2 w}{\partial x \partial y}. \tag{46}$$

Now, a *steady state* response of the half-space is assumed where the displacement varies in the following time-harmonic manner

$$w(x,y,t) = w(x,y) \cdot e^{-i\omega t} \ . \tag{47}$$

In this case, the equation of motion (9) becomes

$$\ell^2 \nabla^4 w - g \nabla^2 w - k^2 w = 0 \ , \tag{48}$$

where $\nabla^2( \ ) = \partial_x^2( \ ) + \partial_y^2( \ )$ is the 2D Laplace operator in Cartesian coordinates, and $(k,g)$ are defined in Section 3.1.

The pertinent boundary conditions for a *traction-free* half-space follow from Eqs. (7) and (8). In particular, for a boundary defined by the plane $y = 0$ with $\mathbf{n} = (0, -1, 0)$, they take the following form

$$P_z^{(n)}(x, y=0) = \tau_{yz} - \frac{\partial m_{xyz}}{\partial x} - \frac{\partial m_{yyz}}{\partial y} - \frac{\partial m_{yxz}}{\partial x} - I\omega^2 \frac{\partial w}{\partial y} = 0 \ , \tag{49}$$

$$R_z^{(n)}(x, y=0) = m_{yyz} = 0 \ . \tag{50}$$

Finally, employing the constitutive equations (45) and (46), we may write the boundary conditions in terms of the displacement $w$ as

$$\ell^2 \frac{\partial^3 w}{\partial y^3} + (2\ell^2 - \bar{a}_3) \frac{\partial^3 w}{\partial x^2 \partial y} - g \frac{\partial w}{\partial y} = 0 \quad \text{for} \quad y=0 \quad \text{and} \quad -\infty < x < \infty \ , \tag{51}$$

$$\ell^2 \frac{\partial^2 w}{\partial y^2} + \bar{a}_3 \frac{\partial^2 w}{\partial x^2} = 0 \quad \text{for} \quad y=0 \quad \text{and} \quad -\infty < x < \infty . \tag{52}$$

*4.2. Integral-transform analysis and the dispersion equation*

In order to suppress the $x$-dependence in the governing equations and the boundary conditions, the Fourier transform is employed. The direct Fourier transform and its inverse are defined as follows (Davies, 2002)

$$f^*(\xi,y)=\int_{-\infty}^{\infty} f(x,y)e^{ix\xi}dx \quad , \quad f(x,y)=\frac{1}{2\pi}\int_{-\infty}^{\infty} f^*(\xi,y)e^{-ix\xi}d\xi \ . \tag{53}$$

Transforming the governing equation (48) with (52)$_1$ gives the following ODE

$$\ell^2 \frac{d^4 w^*}{dy^4} - \left(2\ell^2 \xi^2 + g\right)\frac{d^2 w^*}{dy^2} + \left(\ell^2 \xi^4 + g\xi^2 - k^2\right)w^* = 0 \ . \tag{54}$$

The general transformed solution of (54) has the following bounded at $y \to +\infty$ form

$$w^*(\xi,y) = B(\xi)e^{-\beta y} + C(\xi)e^{-\gamma y} \quad \text{for} \quad y \geq 0 \ , \tag{55}$$

where the functions $(\beta,\gamma)$ are defined in (35) and (36), respectively, and the unknown amplitudes $(B,C)$ can be determined through the enforcement of the pertinent boundary conditions. We note that the branch cuts in Figure 2 are introduced in the complex $\xi$-plane in such a manner that a bounded solution at $y \to +\infty$ is secured. Therefore, any inversion according to (53)$_2$ should be performed considering this restriction (i.e. the cut plane).

The appropriate *dispersion equation* is obtained again by enforcing the pertinent boundary conditions along the traction-free half-space surface $y = 0$. In particular, it is found that the dispersion equation for the motion of progressive SH surface waves in a gradient-elastic homogeneous and isotropic half-space is the same as the one characterizing the propagation of torsional surface waves (c.f. Eq. (40)). More specifically, SH surface waves exist if and only if $\ell^2 \neq 0$ and $\bar{a}_3 \neq 0$; the cases ($\ell^2 = 0$) or ($\ell^2 \neq 0$ and $\bar{a}_3 = 0$) lead to non-existence of SH surface waves. Thus, the simplified version gradient isotropic elasticity ($\bar{a}_3 = 0$) is not capable of predicting such surface waves.

## 5. Numerical results and discussion

In order now to present numerical results in an effective way, the following normalizations are introduced

$$\xi_d = h\xi \quad , \quad \omega_d = \frac{\omega h}{\sqrt{3}c_s} \quad , \quad V_d = \frac{V}{c_s} \quad , \quad \alpha = \frac{\bar{a}_3}{\ell^2} \quad , \quad \varepsilon = \frac{h}{\sqrt{3}\ell} \quad . \tag{56}$$

Note that the normalized parameter $\alpha$ should be bounded by the inequality $|\alpha|<1$ due to positive definiteness of the strain energy, and also $\alpha \neq 0$ for torsional and SH surface waves to exist in a homogeneous and isotropic half-space. Moreover, the wavelength $\lambda$ is introduced through the standard relation $\lambda = 2\pi/\xi$. Three different relations for the ratio of the two microstructural characteristic lengths $h$ and $\ell$ are taken to obtain numerical results, viz. (*i*) $\varepsilon = 2$, (*ii*) $\varepsilon = 1$ and (*iii*) $\varepsilon = 1/2$, whereas six different values of $\alpha$, viz. $\alpha = \{0.1, 0.5, 0.9, 0.999, -0.1, -0.9\}$ are considered in each of the previous cases. All the results presented in this section refer to the propagation of *both* torsional and SH surface waves, since these surface waves are governed by the same dispersion equation in the Toupin-Mindlin theory of gradient elasticity.

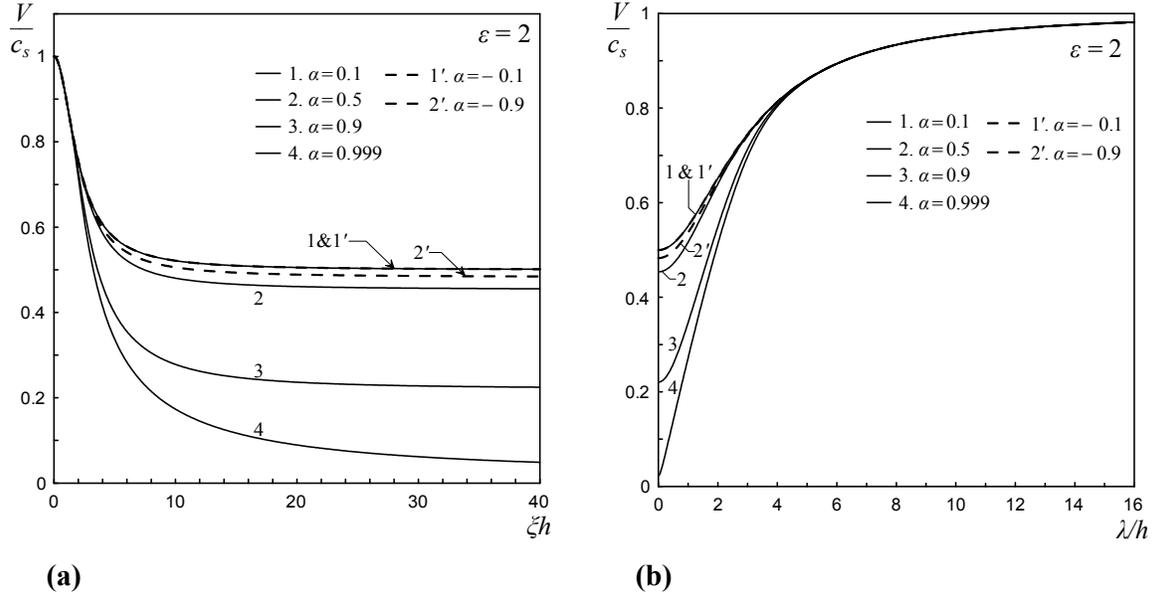

**Fig. 4:** Dispersion curves for the propagation of torsional / SH surface waves showing the variation of the normalized phase velocity $V/c_s$ with the normalized **(a)** wavenumber $\xi h$ and **(b)** wavelength $\lambda/h$, for a microstructured material with $\varepsilon = 2$.

Figures 4a and 4b illustrate the variation of the normalized gradient phase velocity $V_d$ with the normalized wave number $\xi_d$ and the normalized wavelength $\lambda/h$, respectively, for a material with a ratio of the microstructural lengths $\varepsilon = 2$. It is observed that torsional and SH surface waves exist for *all* wavenumbers (no cut-off frequencies appear). This finding is in contrast with previous works by Vardoulakis and Georgiadis (1997), and Georgiadis et al. (2000), where it was shown that torsional and SH waves exist only in certain ranges of frequencies. Moreover, as it is shown in Figure 4a, the phase velocity decreases with increasing wave numbers. This decrease is more pronounced as we approach the upper positive definiteness border $\alpha \to 1$ (curve 4). On the other hand, for increasing negative values of $\alpha$ the decrease of the phase velocity is more moderate. As the wavelength increases (compared to the material microstructure), the phase velocity tends to the classical shear wave velocity $V \to c_S$ (Fig. 4b). This is to be expected intuitively since for relatively long wavelengths the wave should not 'see' the material microstructure. On the other hand, as the wavelength decreases $\lambda/h \to 0$ ($\xi_d \to \infty$), the phase velocity attains a constant value that depends on the microstructural ratios $\varepsilon$ and $\alpha$. This limit value can be analytically evaluated by first noting that the dispersion relation (40) exhibits the following asymptotic behavior as $\xi_d \to \infty$

$$D(\xi,V) \equiv \bar{D}(\xi_d,V_d) = \frac{\left[(1-\alpha)^2 \left(1-\varepsilon^2 V_d^2\right)^{1/2} - \left(1-\alpha-\varepsilon^2 V_d^2\right)^2\right]}{9\sqrt{3}\varepsilon^5} \xi_d^5 + O(\xi_d^3) \ . \qquad (57)$$

The pertinent limit value of the phase velocity of torsional / SH surface waves is given then by the value of $V_d$ for which the coefficient of the leading order term of (57) vanishes. It is noted that, within the bounds of the strain-energy density positive definiteness, this equation has always one real non-zero positive root. As $\alpha \to 1$ or $\varepsilon \to \infty$ the limit velocity becomes zero, while as $\varepsilon \to 0$ (i.e. zero micro-inertia: $h = 0$) the limit velocity becomes infinite as $\xi_d \to \infty$. The last observation can be also inferred from the second equation in (14), where it is apparent that for $h = 0$ the velocity tends to infinity as the wavenumber increases.

Another issue, which merits discussion, pertains to the form of the dispersion curves. In particular, we observe from Figure 4a that $dV/d\xi < 0$ for all wavenumbers, which, in turn, implies that in the case $\varepsilon = 2$, waves exhibit normal dispersion. This is more clearly depicted in Figures 5a and 5b, where the variation of the group velocity $V^g = d\omega/d\xi$ is plotted with respect to the normalized wavenumber. Indeed, it is apparent from Figure 5b that $V > V^g$ for all wavenumbers and, thus, the dispersion is normal, a result in agreement with observations in crystal lattice theories (e.g. Gazis et al., 1960). The difference between the phase velocity and the group velocity becomes more significant for wavenumbers in the range $2 < \xi h < 6$ and as $\alpha \to 1$. For large wavenumbers (high frequencies), the group velocity, in accordance with the behavior of the phase velocity (see Fig. 4), attains a constant value that depends on the material microstructure.

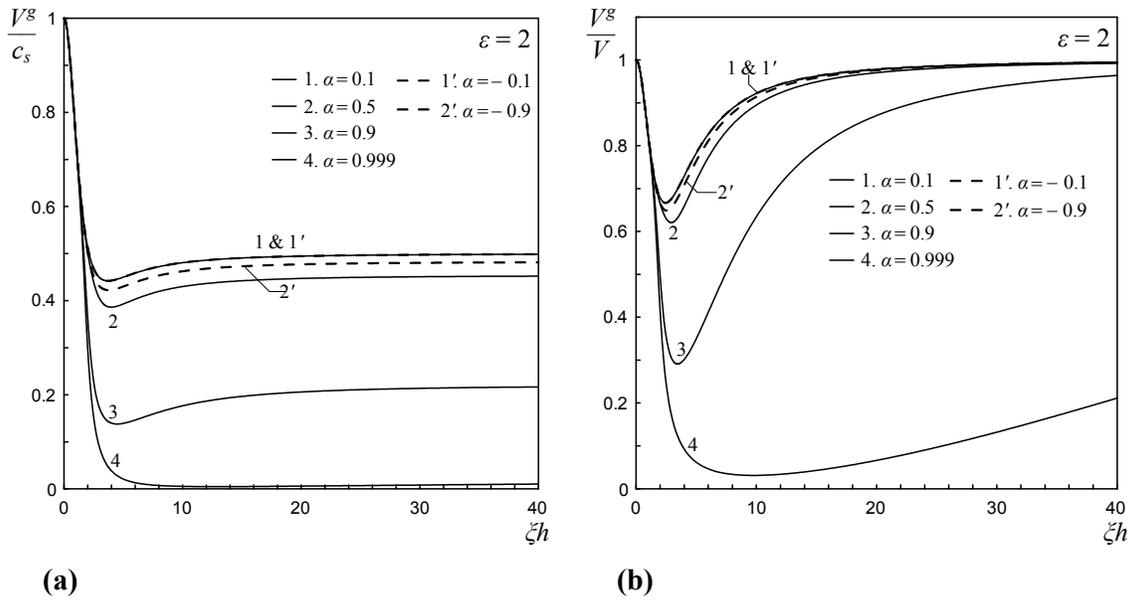

**Fig. 5:** Variation of the group velocity $V^g$ normalized with **(a)** the classical shear wave velocity and **(b)** the gradient phase velocity of the torsional / SH surface waves, versus the normalized wavenumber $\xi h$ for a microstructured material with $\varepsilon = 2$.

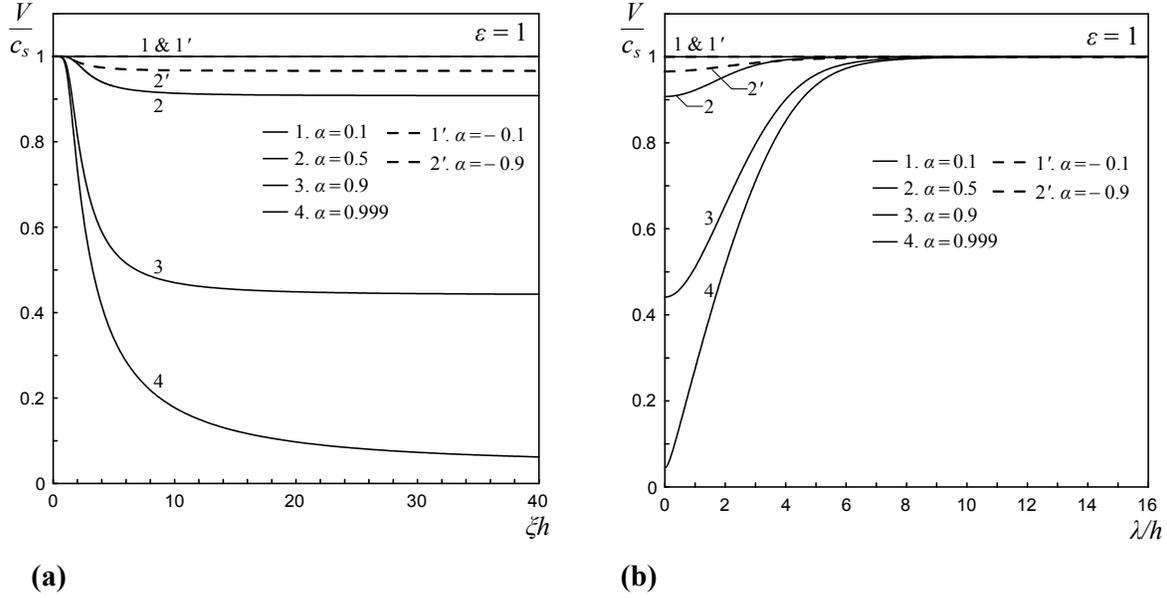

**(a)**                           **(b)**

**Fig. 6:** Dispersion curves for the propagation of torsional / SH surface waves showing the variation of the normalized phase velocity $V/c_s$ with the normalized **(a)** wavenumber $\xi h$ and **(b)** wavelength $\lambda/h$, for a microstructured material with $\varepsilon = 1$.

    Figures 6a and 6b illustrate the variation of the normalized phase velocity $V_d$ with the normalized wave number $\xi_d$ and the normalized wavelength $\lambda/h$, respectively, for a material with $\varepsilon = 1$. It is worth noting that in the case $\varepsilon = 1$, the torsional and SH waves travel without dispersion in an *infinite* medium and their velocities degenerate into the ones of classical elastodynamics (cf. (14)$_2$). For the half-space case considered here, these waves are "almost" non-dispersive in the range $-1 < \alpha < 0.5$ (curves 1, 2, 1′ and 2′ in Figs. 6a and 6b). Indeed, in that range the phase velocity is almost equal to the classical shear wave velocity. However, as $\alpha \to 1$, the dispersive character of these waves becomes more pronounced with increasing wavenumbers (curves 3 and 4). In addition, from Figures 7a and 7b, we observe that $V > V^g$ for all wavenumbers and, thus, the dispersion is again normal. For large wavenumbers (high frequencies), the group velocity, in accordance with the behavior of the phase velocity (see Figs. 6a and 6b), attains a constant value that depends on the microstructural parameters.

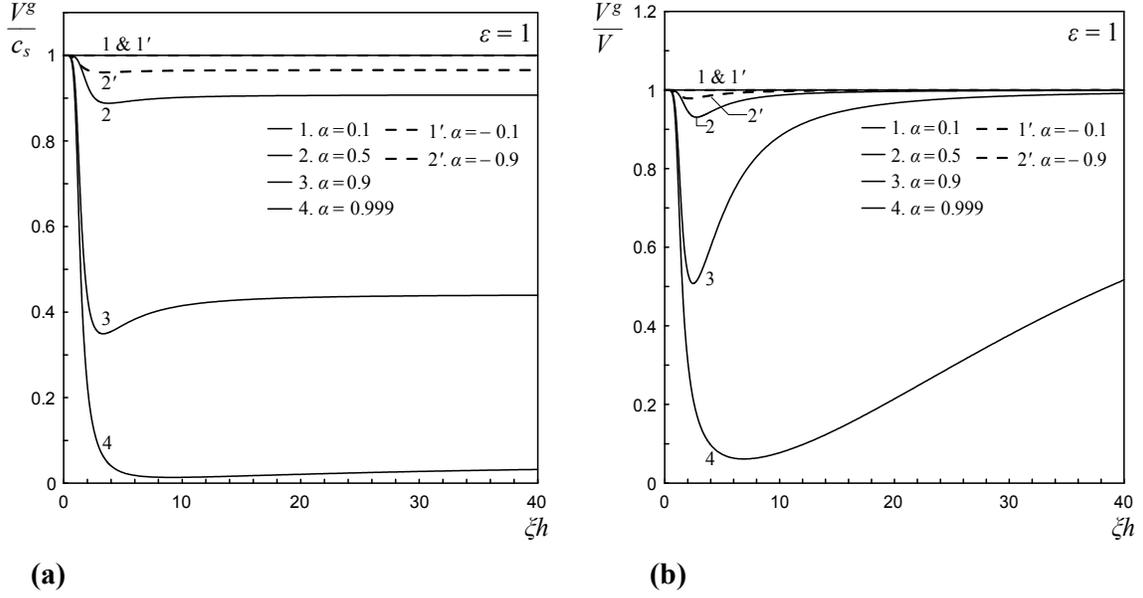

**Fig.7:** Variation of the group velocity $V^g$ normalized with **(a)** the classical shear wave velocity and **(b)** the phase velocity of the torsional / SH surface waves, versus the normalized wavenumber $\xi h$ for a microstructured material with $\varepsilon = 1$.

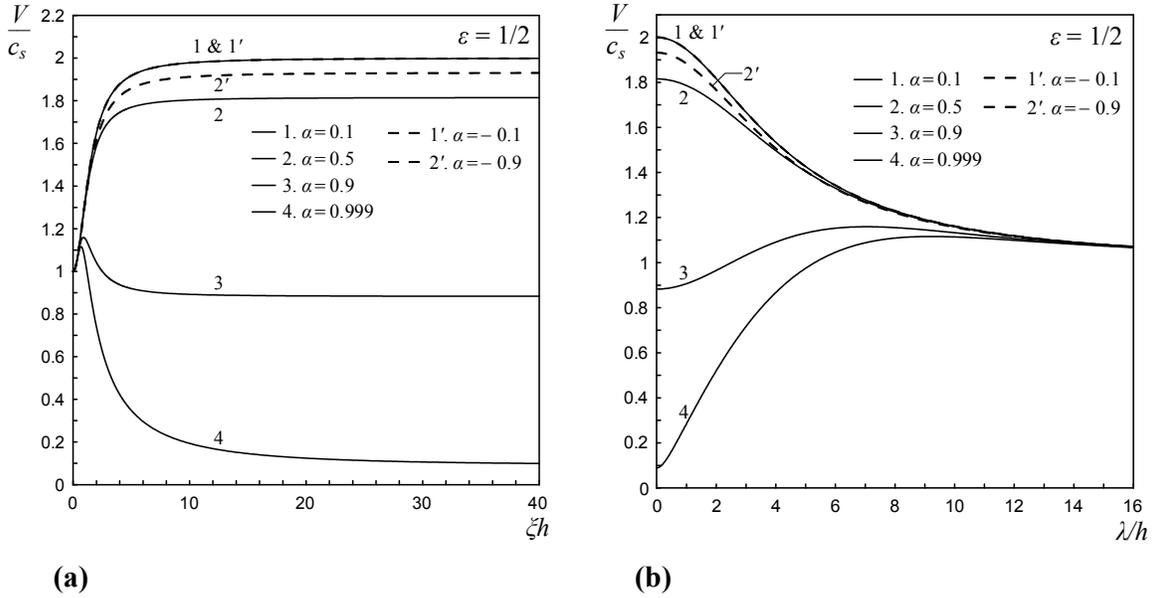

**Fig. 8:** Dispersion curves for the propagation of torsional / SH surface waves showing the variation of the normalized phase velocity $V/c_s$ with the normalized **(a)** wavenumber $\xi h$ and **(b)** wavelength $\lambda/h$, for a microstructured material with $\varepsilon = 1/2$.

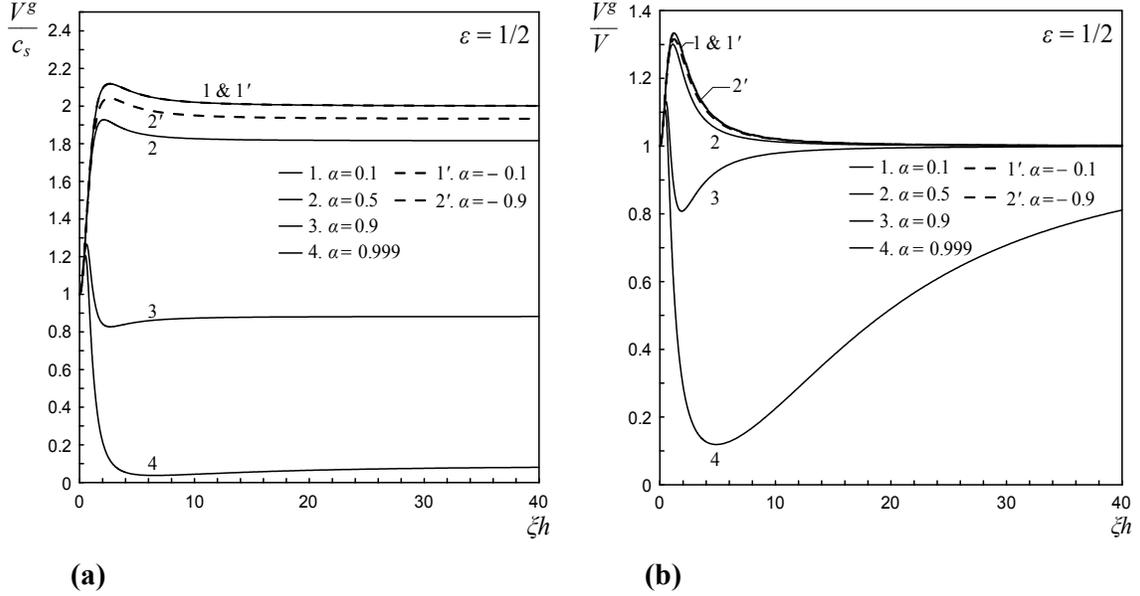

**Fig. 9:** Variation of the group velocity $V^g$ normalized with **(a)** the classical shear wave velocity and **(b)** the phase velocity of the torsional / SH surface waves, versus the normalized wavenumber $\xi h$ for a microstructured material with $\varepsilon = 1/2$.

The case $\varepsilon = 1/2$ is examined in Figures 8 and 9. Contrary to the previous cases (Figs. 4-6), the phase velocity of the torsional and SH waves may exceed now the classical shear wave velocity. More specifically, it is observed from Figure 8a that for small values of the normalized parameter $\alpha$, the normalized phase velocity in gradient elasticity increases above unity with increasing wavenumbers reaching a plateau of a constant limiting value, which can be obtained from Eq. (57). On the other hand, as $\alpha$ increases approaching unity (curves 3 and 4), the phase velocity increases for a small range of wavenumbers ($0 < \xi_d < 2$) and then subsequently decreases resembling the behavior of cases $\varepsilon = 2$ and $\varepsilon = 1$, discussed previously. Regarding the nature of the dispersion curves, we note that for values of the normalized parameter in the range: $-1 < \alpha < 0.8$, the group velocity exceeds the phase velocity $V < V^g$ for all wavenumbers, and thus the dispersion is *anomalous* (Fig. 9b). This finding is agreement with experimental results in granular type materials such as ceramics, sand, concrete, foams, glassy polymers and bones (see e.g. Chen and Lakes, 1989; Giovine and Oliveri, 1995; Stavropoulou et al., 2003; Salupere et al., 2005) Moreover, this behavior reminds analogous results for Stoneley interface waves in a half-space with a superficial layer (Achenbach and Epstein, 1967), and surface waves in liquids that possess surface tension (Coulson, 1958). However, as $\alpha$ increases tending to unity (curves 3 and 4), the dispersion becomes normal again apart from a small initial range of wavenumbers (Fig. 9b).

Finally, Figure 10 shows the variation of the normalized frequency $\omega_d$ of torsional and SH waves with respect to the normalized wavenumber $\xi_d$. It is observed that the form of the dispersion curves depends strongly on the ratio of the microstructural parameters $\varepsilon = h/\sqrt{3}\ell$. In particular, for $\varepsilon \geq 1$ the dispersion is always normal for all wavenumbers. It is worth noting that the same qualitative behavior of the dispersion curves was observed in surface elastic waves and bulk mode vibrations of cubic crystal lattices (Gazis et al., 1960). On the other hand, as $\varepsilon$ decreases ($h \to 0$), the contribution of the micro-inertia is small and an anomalous dispersion behavior prevails. An analogous situation is encountered in many auxetic structures and acoustic metamaterials (see e.g. Chen and Lakes, 1989; Fang et al., 2006)

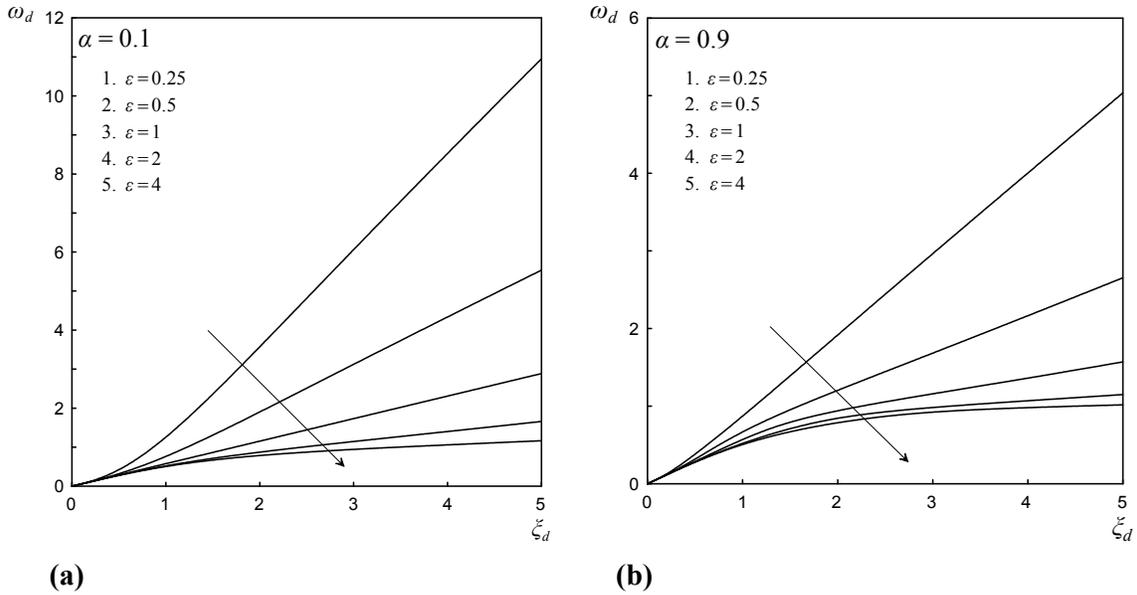

**Fig. 10:** Dispersion curves showing the variation of the normalized frequency $\omega_d$ with the normalized wavenumber $\xi_d$ for a microstructured material with **(a)** $\alpha = 0.1$, **(b)** $\alpha = 0.9$, and various ratios $\varepsilon$.

## 6. Concluding remarks

The present study showed that the existence of torsional and SH surface waves in a *homogeneous* and *isotropic* elastic half-space is possible within the framework of the complete Toupin-Mindlin gradient elasticity theory. The complete theory involves five constants (microstructural parameters) in addition to the standard two Lame moduli. This existence of surface waves is in marked contrast with the well-known result of the classical theory of linear elasticity that torsional / SH surface waves do not exist in a homogeneous half-space. Moreover, our results show that the inclusion of surface energy terms (i.e. gradient anisotropy) used in the past in analogous gradient-type formulations is *not* necessary for the prediction of torsional and SH waves (Vardoulakis et al., 1997; Georgiadis et al., 2000). In particular, it was shown that torsional and SH waves can propagate in a homogeneous and isotropic gradient elastic half-space at *all* frequencies (i.e. no cut-off frequencies appear) and that both waves are governed by the same dispersion equation. In the case where the wavelengths are comparable to the grain size of the material, the dispersion characteristics depend strongly upon the microstructural parameters. In fact, depending on the contribution of the micro-inertia term a normal or an anomalous dispersion behavior may be observed.

## Appendix A

For an *isotropic* gradient material, the positive definiteness of the strain-energy density imposes certain constraints both on the classical and the gradient (microstructural) elastic moduli. Employing a Voigt-type representation, we can rewrite the strain energy density in Eq. (2) as

$$W = \frac{1}{2}\boldsymbol{\varepsilon}^{\mathrm{T}}\mathbf{C}\boldsymbol{\varepsilon} + \frac{1}{2}\boldsymbol{\kappa}^{\mathrm{T}}\mathbf{B}\boldsymbol{\kappa} \ , \tag{A1}$$

where $\mathbf{C}$ is the classical elasticity fourth-order tensor represented here as a $6\times6$ matrix with two independent components $(\lambda,\mu)$ in the isotropic case, and $\mathbf{B}$ is the gradient elasticity fourth-order tensor represented here as an $18\times18$ symmetric matrix with five independent components $(a_1, a_2, a_3, a_4, a_5)$ in the isotropic case. The matrix $\mathbf{B}$ can be written in the following block-diagonal form

$$\mathbf{B} = \begin{bmatrix} \mathbf{D}_1 & \mathbf{0}_1 & \mathbf{0}_1 & \mathbf{0}_2 \\ \mathbf{0}_1 & \mathbf{D}_1 & \mathbf{0}_1 & \mathbf{0}_2 \\ \mathbf{0}_1 & \mathbf{0}_1 & \mathbf{D}_1 & \mathbf{0}_2 \\ \mathbf{0}_2 & \mathbf{0}_2 & \mathbf{0}_2 & \mathbf{D}_2 \end{bmatrix}, \tag{A2}$$

with submatrices

$$\mathbf{D}_1 = \begin{bmatrix} 2(a_2 + a_4) & 2a_2 & a_1 + 2a_5 & a_1 & a_1 + 2a_2 \\ 2a_2 & 2(a_2 + a_4) & a_1 & a_1 + 2a_5 & a_1 + 2a_2 \\ a_1 + 2a_5 & a_1 & 2(a_3 + 2a_4 + a_5) & 2a_3 & a_1 + 2a_3 \\ a_1 & a_1 + 2a_5 & 2a_3 & 2(a_3 + 2a_4 + a_5) & a_1 + 2a_3 \\ a_1 + 2a_2 & a_1 + 2a_2 & a_1 + 2a_3 & a_1 + 2a_3 & 2(a_1 + a_2 + a_3 + a_4 + a_5) \end{bmatrix}, \tag{A3}$$

and

$$\mathbf{D}_2 = 2 \begin{bmatrix} 2a_4 & a_5 & a_5 \\ a_5 & 2a_4 & a_5 \\ a_5 & a_5 & 2a_4 \end{bmatrix}. \tag{A4}$$

The submatrices $\mathbf{0}_1$ and $\mathbf{0}_2$ denote $5 \times 5$ and $3 \times 3$ zero matrices, respectively. In addition, in Eq. (A1), $\boldsymbol{\kappa}$ is a $18 \times 1$ vector with components in the following order

$$\boldsymbol{\kappa} = \{\kappa_{122}, \kappa_{133}, \kappa_{212}, \kappa_{331}, \kappa_{111}, \kappa_{211}, \kappa_{233}, \kappa_{112}, \kappa_{332}, \kappa_{222}, \kappa_{311},$$
$$\kappa_{322}, \kappa_{113}, \kappa_{223}, \kappa_{333}, \kappa_{123}, \kappa_{213}, \kappa_{312}\}. \tag{A5}$$

The determinant of the matrix $\mathbf{B}$ can be represented in the form of a product of four determinants

$$|\mathbf{B}| = |\mathbf{D}_1|^3 |\mathbf{D}_2|. \tag{A6}$$

In accordance with Sylvester's criterion, positive definiteness of the quadratic form in (A1) requires the leading principal minor (PM) determinants of $\mathbf{C}$ and $\mathbf{B}$ to be positive. Regarding the elasticity tensor $\mathbf{C}$, the usual inequalities for the Lamé moduli are obtained: $3\lambda + 2\mu > 0$ and $\mu > 0$. On the other hand, for the gradient elasticity tensor $\mathbf{B}$ the pertinent conditions can be derived by examining separately the determinants of $\mathbf{D}_1$ and $\mathbf{D}_2$. In particular, from the leading PM determinants of $\mathbf{D}_2$, we deduce the following inequalities

$$a_4 > 0, \quad a_4 + a_5 > 0, \quad 2a_4 - a_5 > 0. \tag{A7}$$

Further, the necessary conditions for the matrix $\mathbf{B}$ to be positive definite is that all its diagonal elements are positive, thus

$$a_1 + a_2 + a_3 + a_4 + a_5 > 0, \quad a_3 + 2a_4 + a_5 > 0, \quad a_2 + a_4 > 0. \tag{A8}$$

Multiplying (A8)$_1$ by the factor 2 and adding it to (A7)$_2$, we obtain also that

$$b_0 \equiv 2a_1 + 2a_2 + 2a_3 + 3a_4 + 3a_5 > 0. \tag{A9}$$

The first three leading PM determinants of $\mathbf{D}_1$ do not yield any new inequalities for the parameters $a_j$, whereas the fourth leading PM furnishes

$$(a_4 + a_5)(2a_4 - a_5)\left[(-4a_1 + 8a_2 + 2a_3 + 6a_4 - 3a_5)b_0 - (a_1 + 4a_2 - 2a_3)^2\right] > 0, \tag{A10}$$

which, in turn, implies that

$$b_1 \equiv -4a_1 + 8a_2 + 2a_3 + 6a_4 - 3a_5 > 0. \tag{A11}$$

Moreover, from the requirement that $|\mathbf{D}_1| > 0$ (5$^{\text{th}}$ leading PM), we have that

$$(a_4 + a_5)^2 (2a_4 - a_5)\left[2b_1\left(5(a_1 + a_2 + a_3) + 3(a_4 + a_5)\right) - 5(a_1 + 4a_2 - 2a_3)^2\right] > 0, \tag{A12}$$

which, in view of the above, shows that

$$b_2 \equiv 5(a_1 + a_2 + a_3) + 3(a_4 + a_5) > 0 \quad \text{and} \quad 2b_1 b_2 - 5(a_1 + 4a_2 - 2a_3)^2 > 0 \ . \tag{A13}$$

In addition, combining the inequalities (A13) with (A11) yields

$$10a_3 + 6a_4 + a_5 > 0 \ . \tag{A14}$$

It is worth noting that the inequalities (A7), (A11) and (A13) are equivalent to the ones obtained by Mindlin and Eshel (1968), and Eshel and Rosenfeld (1970) in the Form III of Mindlin's gradient elasticity theory (Mindlin, 1964).

Finally, recalling that $a_3 + 2a_4 + a_5 = 2\mu \ell^2$, we derive, according to (A7)$_2$, (A8)$_2$ and (A14), the following bounds of positive definiteness for the parameter $a_3$

$$-\ell^2 < \bar{a}_3 < \ell^2 \ , \tag{A15}$$

with $\bar{a}_3 = a_3 / 2\mu$.

**Acknowledgement**

Panos A. Gourgiotis gratefully acknowledges support from the ERC Advanced Grant 'Instabilities and nonlocal multiscale modelling of materials' FP7-PEOPLE-IDEAS-ERC-2013-AdG (2014-2019).

**References**

Achenbach, J.D., 1984. Wave propagation in elastic solids. North-Holland, Amsterdam.

Achenbach, J.D., Balogun, O., 2010. Anti-plane surface waves on a half-space with depth-dependent properties. *Wave Motion* 47, 59-65.

Achenbach, J.D., Epstein, H.I., 1967. Dynamic interaction of a layer and a half-space. *J. Eng. Mech.* 5, 27-42.

Bleustein, J.L., 1967. A note on the boundary conditions of Toupin's strain-gradient theory. *Int. J. Solids Struct.* 3, 1053-1057.

Bullen, K.E., Bolt, B.A., 1985. An introduction to the theory of seismology. Cambridge University Press, London.

Chattaraj, R., Samal, S.K., Mahanti, N.C., 2011. Propagation of torsional surface wave in anisotropic poroelastic medium under initial stress. *Wave Motion* 48, 184-195.

Chen, C.P., Lakes, R.S., 1989. Dynamic wave dispersion and loss properties of conventional and negative poisson's ratio polymeric cellular materials. *Cell. Polym.* 8, 343-359.

Collet, B., Destrade, M., Maugin, G.A., 2006. Bleustein–Gulyaev waves in some functionally graded materials. *Eur. J. Mech. A-Solid* 25, 695-706.

Coulson, C.A., 1958. Waves. Oliver and Boyd, Edinburgh.

Davies, B., 2002. Integral transforms and their applications. Springer, New York.

Dey, S., Gupta, S., Gupta, A., 1993. Torsional surface wave in an elastic half-space with void pores. *Int. J. Numer. Anal. Methods Geomech.* 17, 197-204.

Du, C., Su, X., 2013. SH surface waves in a half space with random heterogeneities, Computational methods in stochastic dynamics. Springer, pp. 255-266.

Eringen, A.C., 1972. Linear theory of nonlocal elasticity and dispersion of plane waves. *Int. J. Eng. Sci.* 10, 425-435.

Eringen, A.C., Suhubi, E.S. 1975. Elastodynamics, Vol. 2. Academic Press, New York.

Eshel, N.N., Rosenfeld, G., 1970. Effects of strain-gradient on the stress-concentration at a cylindrical hole in a field of uniaxial tension. *J. Eng. Math.* 4, 97-111.

Fafalis, D.A., Filopoulos, S.P., Tsamasphyros, G.J., 2012. On the capability of generalized continuum theories to capture dispersion characteristics at the atomic scale. *Eur. J. Mech. A-Solid* 36, 25-37.

Fang, N., Xi, D., Xu, J., Ambati, M., Srituravanich, W., Sun, C., Zhang, X., 2006. Ultrasonic metamaterials with negative modulus. *Nat. Mater.* 5, 452-456.

Filopoulos, S., Papathanasiou, T.K., Markolefas, S.I., Tsamasphyros, G.J., 2010. Dynamic finite element analysis of a gradient elastic bar with micro-inertia. *Comput. Mech.* 45, 311-319.

Gao, X.-L., Ma, H., 2009. Green's function and Eshelby's tensor based on a simplified strain gradient elasticity theory. *Acta Mech.* 207, 163-181.

Gazis, D.C., Herman, R., Wallis, R.F., 1960. Surface elastic waves in cubic crystals. *Physical Review* 119, 533.

Georgiadis, H.G., Anagnostou, D.S., 2008. Problems of the Flamant–Boussinesq and Kelvin type in dipolar gradient elasticity. *J. Elast.* 90, 71-98.

Georgiadis, H.G., Vardoulakis, I., Lykotrafitis, G., 2000. Torsional surface waves in a gradient-elastic half-space. *Wave Motion* 31, 333-348.

Georgiadis, H.G., Vardoulakis, I., Velgaki, E.G., 2004. Dispersive rayleigh-wave propagation in microstructured solids characterized by dipolar gradient elasticity. *J. Elast.* 74, 17-45.

Georgiadis, H.G., Velgaki, E.G., 2003. High-frequency rayleigh waves in materials with micro-structure and couple-stress effects. *Int. J. Solids Struct.* 40, 2501-2520.

Giannakopoulos, A.E., Petridis, S., Sophianopoulos, D.S., 2012. Dipolar gradient elasticity of cables. *Int. J. Solids Struct*. 49, 1259-1265.

Giovine, P., Oliveri, F., 1995. Dynamics and wave propagation in dilatant granular materials. *Meccanica* 30, 341-357.

Gourgiotis, P.A., Georgiadis, H.G., 2009. Plane-strain crack problems in microstructured solids governed by dipolar gradient elasticity. *J. Mech. Phys. Solids* 57, 1898-1920.

Gourgiotis, P.A., Georgiadis, H.G., Neocleous, I., 2013. On the reflection of waves in half-spaces of microstructured materials governed by dipolar gradient elasticity. *Wave Motion* 50, 437-455.

Grentzelou, C.G., Georgiadis, H.G., 2008. Balance laws and energy release rates for cracks in dipolar gradient elasticity. *Int. J. Solids Struct.* 45, 551-567.

Jaunzemis, W., 1967. Continuum mechanics. Macmillan, New York.

Kraut, E., 1971. Surface elastic waves - a review. *Acoustic Surface Wave and Acousto-optic Devises,* Optosonic Press, New York.

Maugin, G., 1988. Shear horizontal surface acoustic waves on solids, *Recent developments in surface acoustic waves*. Springer, pp. 158-172.

Meissner, E., 1921. Elastische oberflachenwellen mit dispersion in einem inhomogenen medium. *Vierteljahrsschrift der Naturforschenden Gesellschaft in Zurich* 66, 181-195.

Mindlin, R.D., 1964. Micro-structure in linear elasticity. *Arch. Ration. Mech. Anal.* 16, 51-78.

Mindlin, R.D., Eshel, N.N., 1968. On first strain-gradient theories in linear elasticity. *Int. J. Solids Struct.* 4, 109-124.

Morini, L., Piccolroaz, A., Mishuris, G., 2014. Remarks on the energy release rate for an antiplane moving crack in couple stress elasticity. *Int. J. Solids Struct.* 51, 3087-3100.

Piccolroaz, A., Movchan, A.B., 2014. Dispersion and localisation in structured rayleigh beams. *Int. J. Solids Struct*. 51, 4452-4461.

Polyzos, D., Fotiadis, D.I., 2012. Derivation of Mindlin's first and second strain gradient elastic theory via simple lattice and continuum models. *Int. J. Solids Struct.* 49, 470-480.

Rayleigh, L., 1885. On waves propagated along the plane surface of an elastic solid. *Proc. R. Soc. London, Sect. A.* 17, 4-11.

Rosi, G., Nguyen, V.-H., Naili, S., 2014. Reflection of acoustic wave at the interface of a fluid-loaded dipolargradient elastic half-space. *Mech. Res. Commun.* 56, 98-103.

Salupere, A., Engelbrecht, J., Ilison, O., Ilison, L., 2005. On solitons in microstructured solids and granular materials. *Math. Comput. Simul* 69, 502-513.

Shodja, H.M., Zaheri, A., Tehranchi, A., 2013. Ab initio calculations of characteristic lengths of crystalline materials in first strain gradient elasticity. *Mech. Mater.* 61, 73-78.

Shuvalov, A.L., Poncelet, O., Golkin, S.V., 2009. Existence and spectral properties of shear horizontal surface acoustic waves in vertically periodic half-spaces. *Proc. R. Soc. London, Ser. A* 465, 1489-1511.

Thompson, J., 1969. Some existence theorems for the traction boundary value problem of linearized elastostatics. *Arch. Ration. Mech. Anal.* 32, 369-399.

Ting, T.C.T., 2010. Existence of anti-plane shear surface waves in anisotropic elastic half-space with depth-dependent material properties. *Wave Motion* 47, 350-357.

Toupin, R.A., 1962. Elastic materials with couple-stresses. *Arch. Ration. Mech. Anal.* 11, 385-414.

Vardoulakis, I., 1984. Torsional surface waves in inhomogeneous elastic media. *Int. J. Numer. Anal. Methods Geomech.* 8, 287-296.

Vardoulakis, I., Georgiadis, H.G., 1997. SH surface waves in a homogeneous gradient-elastic half-space with surface energy. *J. Elast.* 47, 147-165.

Vavva, M.G., Protopappas, V.C., Gergidis, L.N., Charalambopoulos, A., Fotiadis, D.I., Polyzos, D., 2009. Velocity dispersion of guided waves propagating in a free gradient elastic plate: Application to cortical bone. *J. Acoust. Soc. Am.* 125, 3414-3427.